\documentstyle[emulateapj]{article}
\newcommand{\co}{\rm CO}
\newcommand{\hh}{{\rm H}_2}

\newcommand{\kps}{\,\textstyle\rm{km~s}^{-1}}

\newcommand{\Mpc}{\,\textstyle\rm{Mpc}}
\newcommand{\coc}{CO(3$\rightarrow$2) }

\newcommand{\yr}{\,\textstyle\rm{yr}}

\newcommand{\msun}{\,M_{\odot}}
\newcommand{\lsun}{\,L_{\odot}}

\newcommand{\erg}{\,\textstyle\rm{erg}}
\newcommand{\Hz}{\,\textstyle\rm{Hz}}
\newcommand{\s}{\,\textstyle\rm{s}}

\newcommand{\jy}{\,\textstyle\rm{Jy}}

\newcommand{\Kkpspc}{\,\rm{K}\,\rm{km~s}^{-1}\,{\rm pc}^{2}}

\lefthead{Frayer et al.}
\righthead{Molecular Gas in SMM\,J14011+0252}

\begin{document}

\title{Molecular Gas in the {\scriptsize z}=2.565 Submillimeter Galaxy
SMM\,J14011+0252}

\author{D. T. Frayer\altaffilmark{1}, R. J. Ivison\altaffilmark{2},
N. Z. Scoville\altaffilmark{1}, A. S. Evans\altaffilmark{1}, 
M. Yun\altaffilmark{3}, Ian Smail \altaffilmark{4},
A. J. Barger\altaffilmark{5}, A. W. Blain\altaffilmark{6}, and
J.--P. Kneib\altaffilmark{7}}

\altaffiltext{1}{Astronomy Department, California Institute of Technology
105--24, Pasadena, CA  91125, USA} 

\altaffiltext{2}{Department of Physics \& Astronomy, University College
London, Gower Street, London, WC1E 6BT, UK}

\altaffiltext{3}{National Radio Astronomy Observatory, P.O. Box 0,
Socorro, NM  87801, USA}

\altaffiltext{4}{Department of Physics, University of Durham, South
Road, Durham, DH1 3LE, UK}

\altaffiltext{5}{Institute for Astronomy, University of Hawaii, 2680
Woodlawn Drive, Honolulu, HI 96822}

\altaffiltext{6}{Cavendish Laboratory, Madingley Road, Cambridge, CB3 OHE, UK}

\altaffiltext{7}{Observatoire Midi--Pyr\'{e}n\'{e}es, 14 Avenue E. Belin,
F--31400 Toulouse, France}

\begin{abstract}

We report the detection of \coc emission from the
submillimeter--selected luminous galaxy SMM\,J14011+0252.  The optical
counterpart of the submillimeter source has been identified as a merger
system with spectral characteristics consistent with a starburst at
$z=2.565$.  The CO emission confirms the optical identification of the
submillimeter source and implies a molecular gas mass of $5\times
10^{10} h_{75}^{-2} M_{\sun}$, after correcting for a lensing
amplification factor of 2.75.  The large molecular gas mass and the
radio emission are consistent with the starburst interpretation of the
source.  These results are similar to those found for SMM\,J02399--0136,
which was the first submillimeter selected CO source found at high
redshift.  The CO detections of these two high--redshift submillimeter
galaxies suggest the presence of massive reservoirs of molecular gas
which is consistent with the inferred high rates of star--formation
($10^{3}\,M_{\sun}\yr^{-1}$).  These two systems appear to be associated
with merger events which may evolve into present day luminous elliptical
galaxies.

\end{abstract}


\section{INTRODUCTION}

Deep surveys of the submillimeter sky using the Submillimeter
Common--User Bolometer Array (SCUBA) on the James Clerk Maxwell
Telescope have uncovered a population of distant dust--rich galaxies
(Smail, Ivison \& Blain 1997; Blain et al.\ 1998; Barger et al.\ 1998;
Hughes et al.\ 1998; Eales et al.\ 1998).  Based on their optical
colors, the majority of these systems are thought to lie at redshifts of
$z\sim 1$--5 (Smail et al.\ 1998b).  However, only a few of these
systems currently have accurate redshifts.  The first well--studied
system SMM\,J02399--0136, hereafter SMM\,J02399, at $z=2.8$ was shown to
contain both an AGN (Ivison et al.\ 1998) and a massive reservoir of
molecular gas thought to be fueling a starburst (Frayer et al.\ 1998).

In order to test the generality of the properties of SMM\,J02399, we
have initiated a search for CO emission in additional sub-mm sources.
In this {\em Letter} we report the results for the submillimeter source
SMM\,J14011+0252, hereafter SMM\,J14011.  This sub-mm galaxy was
discovered during a survey through rich, lensing clusters (Smail et al.\
1998b).  Follow-up optical spectroscopy places the sub-mm source at a
redshift of approximately $z=2.55$ (Barger et al.\ 1999; Ivison et al.\
1999).  SMM\,J14011 appears to be a merger/interacting system which has
optical and radio properties consistent with an ultraluminous starburst
(Ivison et al.\ 1999), albeit the presence of a heavily obscured AGN
cannot be ruled out.  The amplification factor for SMM\,J14011 is
$2.75\pm0.25$, where the uncertainty comes from the allowed range of
detailed mass models of the foreground cluster.

\section{OBSERVATIONS}

SMM\,J14011 was observed using the Owens Valley Millimeter Array between
1998 October and December.  A total of 41 hours of usable
integration time on source was obtained in two configurations of six
10.4m telescopes.  The phase center for the CO observations was the
position of the brightest optical component of the SMM\,J14011 system
(J1); $\alpha$(J2000)=$14^{\rm h} 01^{\rm m}04\fs97$;
$\delta$(J2000)=$+02\arcdeg 52\arcmin 24\farcs6$ (Ivison et al.\ 1999).
The \coc line was observed using a digital correlator configured with
$112\times4$~MHz channels centered on 96.9975 GHz in the lower
side--band, corresponding to \coc emission at the H$\alpha$ redshift of
$z=2.565$ (Ivison et al.\ 1999).  Typical single--sideband system
temperatures were approximately 300--350 K, corrected for telescope
losses and the atmosphere.  In addition to the CO line data, we recorded
the 3mm continuum data with a 1 GHz bandwidth for both the upper
(line--free, centered on 99.9975 GHz) and lower sidebands.  The nearby
quasar 1413+135 was observed every 25 minutes for gain and phase
calibration.  Absolute flux calibration was determined from observations
of Uranus, Neptune, and 3C\,273.  The absolute calibration uncertainty
for the data is approximately 15\%.

\section{RESULTS}

Figure 1 shows the \coc spectrum for SMM\,J14011.  The CO line is
detected at the redshift of the narrow H$\alpha$ line of SMM\,J14011
(Ivison et al.\ 1999).  We achieved a $7\sigma$ detection for the peak
in the integrated \coc map (Fig. 2).  The CO position (Table~1) is
consistent with the optical position of SMM\,J14011 within the
uncertainties of the data sets.  The CO map shows marginal evidence for
emission extended toward the south, but observations at higher
resolution are required to test this possibility.  The upper--limit of
the continuum emission is S$_{\nu}$(3\,mm) $< 0.6$\,mJy ($1\sigma$).
This is insufficient to detect the thermal dust emission discovered by
SCUBA, assuming $S_{\nu} \propto \nu^{3.5}$.

Both the CO and H$\alpha$ lines of SMM\,J14011 are redshifted from the
UV lines by  about $1100\pm700\kps$ (Ivison et al.\ 1999).  Systematic
blue--ward offsets of UV lines are not unusual for starbursts and have
been attributed to outflows with dust obscuration at systemic and
redshifted velocities (Mirabel \& Sanders 1988; Gonz\'{a}lez Delgado et
al.\ 1998; Heckman et al.\ 1998).  In an extreme example, the $z=3.9$
quasar APM\,08279+5255 shows a 2500$\kps$ blue--ward offset of the UV
lines from the systemic CO redshift (Downes et al.\ 1998).

The observed \coc line flux is $S(\co) = 2.4\pm0.3 \jy\kps$.  No
adjustment has been made to account for the continuum level since it is
negligible.  The observed \coc line flux implies an intrinsic CO line
luminosity\footnote{We adopt $q_o=0.5$, $H_o=75\,h_{75} \kps\Mpc^{-1}$,
and a lensing amplification factor of 2.75 throughout this {\em
Letter}.} of $L^{\prime}(\co) = 1.2\times 10^{10} h_{75}^{-2} \Kkpspc$
(see formulae in Solomon, Downes, \& Radford 1992).  The CO luminosity
is related to the mass of molecular gas (including He) by
$M(\hh)/L^{\prime}(\co) = \alpha$.  The value for $\alpha$ is expected
to be between $\alpha \simeq 1 \msun(\Kkpspc)^{-1}$ (Solomon et al.\
1997) and the Galactic value of $\alpha \simeq 5 \msun(\Kkpspc)^{-1}$
(e.g., Sanders, Scoville, \& Soifer 1991).  We adopt a value of
$\alpha=4 \msun(\Kkpspc)^{-1}$, which is consistent with that estimated
for Arp~220 (Scoville, Yun, \& Bryant 1997a) after correcting for the
line brightness ratio of
$T_b(\co[3\rightarrow2])/T_b(\co[1\rightarrow0])\simeq 0.6$ typically
observed in starbursts (Devereux et al.\ 1994).  The inferred molecular
gas mass of SMM\,J14011 is $5 \times 10^{10} h_{75}^{-2} M_{\sun}$,
which is consistent with that of the most massive low--redshift
ultraluminous infrared galaxies (ULIGs; Sanders \& Mirabel 1996).

The implied gas--to--dust ratio for SMM\,J14011 is $M(\hh)/M{\rm
dust}=$\,200--900, assuming a reasonable range of possible dust
temperatures (30--70\,K).  These gas--to--dust ratios are similar to
those seen in spiral galaxies (Devereux \& Young 1990), local ULIGs
(Sanders et al.\ 1991), and other high--redshift CO sources.  The
similarities of the derived gas--to--dust ratios at high--redshift
reflect the small scatter (only a factor of 2--3) in their
$L^{\prime}(\co)/L_{\nu}({\rm submm})$ luminosity ratios.  These results
are somewhat remarkable considering that some of the sources are more
likely dominated by AGNs than others and considering the possibility of
different dust temperatures.

The observed line width for SMM\,J14011 (FWHM=$200\kps$) is much
narrower than that observed for SMM\,J02399 ($700\kps$), but is within
the range of values typically observed for ULIGs (200--$400\kps$;
Solomon et al.\ 1997).  The narrow line--width may indicate that we are
viewing the system somewhat face-on.  By using the observed line width
and the upper limit on the CO source size of $\theta < 7\arcsec$, we can
constrain the total dynamical mass contained within the CO emission
regions.  The angular size limit corresponds to a maximum linear
diameter of $D < 36 h_{75}^{-1}$ kpc at $z=2.565$, or $D < 13 h_{75}^{-1}$
kpc after correcting for lensing.  The dynamical mass is $M_{dyn}\simeq
R(\Delta V/[2\sin(i)])^2/G$, where $\Delta V$ is the observed FWHM line
width and $i$ is the inclination.  We find $M_{dyn} < 1.5 \times 10^{10}
\sin^{-2}(i)\,h_{75}^{-1} M_{\sun}$ which is consistent with the derived
gas mass, provided $i \la 35 \arcdeg$.

\section{DISCUSSION}

The intrinsic far--infrared luminosity of SMM\,J14011 is approximately
$3\times 10^{12} h_{75}^{-2}\lsun$.  If its luminosity is powered by
star--formation, as suggested by the optical data (Ivison et al.\ 1999),
the implied star--formation rate of massive stars ($M>5\msun$) is
$300\msun\yr^{-1}$ (Condon 1992).  Including the presence of low mass
stars, the total star formation rate would be of order
$10^{3}\msun\yr^{-1}$.  The CO data is consistent with this view by
showing the presence of enough gaseous material to support a high level
of star formation.  The large molecular gas mass and the high
star-formation rate suggest a starburst time scale of order
$5\times10^{7} \yr$.  In addition, the far-infrared--to--radio flux
ratio for SMM\,J14011 is the same as that found for nearby starburst
galaxies (Condon, Frayer, \& Broderick 1991).  Alternatively, if
SMM\,J14011 is powered by an AGN, the AGN could not be a powerful radio
source ($<3\times10^{30}h_{75}^{-2}\erg\s^{-1}\Hz^{-1}$).

The success of the first two searches for CO from the high--redshift
submillimeter--selected galaxies is encouraging.  Unlike SMM\,J02399,
whose UV/optical spectral properties show significant AGN activity,
there is currently no evidence for an AGN in SMM\,J14011.  However, the
presence of a dust--enshrouded AGN cannot be completely ruled out.  Even
though the UV/optical characteristics of SMM\,J14011 and SMM\,J02399
appear different, their radio, submillimeter, and CO properties are
fairly similar and suggest that star formation is important in the
far--infrared emission of both sources.  SMM\,J14011 and SMM\,J02399
share many of the same properties of the local population of ULIGs, such
as (1) high infrared luminosities, (2) associated with mergers, (3)
large reservoirs of molecular gas, and (4) comparable CO line widths.
The most important distinction is that the high-redshift sub-mm galaxies
are about a factor of $10^2$--$10^{3}$ times more numerous per unit
co--moving volume than the low redshift ULIGs (Kim \& Sanders 1998).
The sub-mm galaxies may also have slightly larger amounts of molecular
gas (by factors of 2--3) than typical ULIGs.

Figure~3 shows the CO luminosities plotted as a function of redshift.
When lensing corrections are taken into account, the high--redshift
sources have similar CO luminosities to the most luminous low redshift
ULIGs.  These consistencies could suggest roughly similar gas masses in
the progenitor systems or may indicate the importance of
self--regulating mechanisms such as galactic winds (Heckman, Armus, \&
Miley 1990).  For comparison, theoretical evolutionary curves are also
presented in Figure~3.  We show three models based on numerical
calculations of chemical evolution (Frayer \& Brown 1997).  The models
are designed to represent a spiral galaxy, an elliptical galaxy, and a
merger event.  All three models start as purely gaseous systems with
zero metallicity.  The spiral galaxy model uses an infall model (Run~11
of Frayer \& Brown 1997) which starts at $z=5$ and reaches a final mass
similar to that of the current Milky Way disk ($6\times10^{10}\msun$).
The CO luminosity of a disk with the mass of the Milky Way is never
expected to reach the high CO luminosities ($\ga 10^{10} \Kkpspc$)
observed in the high-redshift systems.  For the elliptical galaxy model,
we assume a closed-box model (Run~8 of Frayer \& Brown 1997) with a mass
of $2\times 10^{11}\msun$ which begins its evolution at $z=5$.  Although
galactic systems are not expected to evolve in such a simple manner, the
closed-box model is able to match the high CO luminosities detected at
high redshift and the low CO luminosities ($\la 10^{7}\Kkpspc$) observed
for local elliptical galaxies (Wiklind, Combes, \& Henkel 1995).  The
third model in Figure~3 is a more realistic model for SMM\,J14011 and
SMM\,J02399.  It assumes a starburst resulting from the merger of two
$2\times10^{11}\msun$ gaseous systems at $z=3$.  This starburst model
uses the parameters of Run~8 in Frayer \& Brown (1997), except for a
factor of 10 increase in the star formation efficiency and assumes a
formation time scale of $\tau_f=3\times10^{8}\yr$ for the merger (Mihos
\& Hernquist 1996).  These theoretical calculations only show the
approximate evolution expected for different types of galactic systems
and depend strongly on the parameters governing star formation, the
initial mass function, infall, and cosmology.

An important unanswered question is what will SMM\,J14011 evolve into at
the present epoch.  By using the observed near--infrared K-band flux
density (Ivison et al.\ 1999) and assuming the burst model from Bruzual
\& Charlot (1993), SMM\,J14011 has already formed of order
$10^{10}\msun$ of stars.  There is enough molecular gas present to form
an additional $L^{*}$ galaxy (adopting $5\times10^{10}\msun$ per
$L^{*}$), suggesting that SMM\,J14011 will evolve into a $\sim
1$--$2\,h_{75}^{-2}\,L^{*}$ galaxy.  Given that the low--redshift ULIGs
may eventually evolve into elliptical galaxies (Kormendy \& Sanders
1992; Mihos \& Hernquist 1996), the high-redshift sub-mm population
could represent luminous elliptical galaxies in their formative phases
(Smail et al.\ 1997; Eales et al.\ 1998).  The consistency of this
hypothesis can be checked by comparing the number density of the sub-mm
population represented by SMM\,J14011 and SMM\,J02399 with that of
luminous ellipticals at low redshift.  Within an effective source plane
area of 25 square arcminutes, Smail et al.\ (1998b) have detected 9
distant, non-cluster, sub-mm sources above the $4\sigma$ noise level.
SMM\,J14011 and SMM\,J02399 represent about 20\% of this sample.  The
corresponding number density of such sources per unit co-moving volume
within the redshift range of $z=1$--5 is approximately $10^{-4}
h_{75}^{3}\Mpc^{-3}$.  The number density of luminous elliptical
galaxies found at low redshift is roughly similar (Loveday et al.\
1992).  Although no definitive conclusions can be made due to the small
sample size, the current data are at least consistent with the sub-mm
population evolving into the luminous elliptical galaxies at the present
epoch.

\section{CONCLUSIONS}

We report the detection of CO in a second ultraluminous galaxy,
SMM\,J14011, selected from a submillimeter survey of the distant
Universe (Smail et al.\ 1998b).  The CO emission is coincident in
position and redshift with the optical counterpart of SMM\,J14011 and
indicates the presence of approximately $5\times 10^{10} h^{-2}_{75}
M_{\sun}$ of molecular gas.  Both SMM\,J14011 and the first
high-redshift sub-mm galaxy, SMM\,J02399, show CO emission and are
thought to be associated with a major burst of star formation.  The
massive molecular gas reservoirs of SMM\,J14011 and SMM\,J02399 and
their co-moving number densities are consistent with these sources
evolving into present day luminous elliptical galaxies.  Further
observations are needed to constrain the redshift distribution and the
molecular gas masses of sub-mm population in order to test the
generality of these early results.

\acknowledgments

We thank our colleagues at the Owens Valley Millimeter Array who have
helped make these observations possible.  The Owens Valley Millimeter
Array is operated by the California Institute of Technology and is
supported by NSF grants AST 93--14079 and AST 96--13717.  RJI
acknowledges a PPARC Advanced Fellowship and IRS acknowledges a Royal
Society Fellowship.

\figcaption[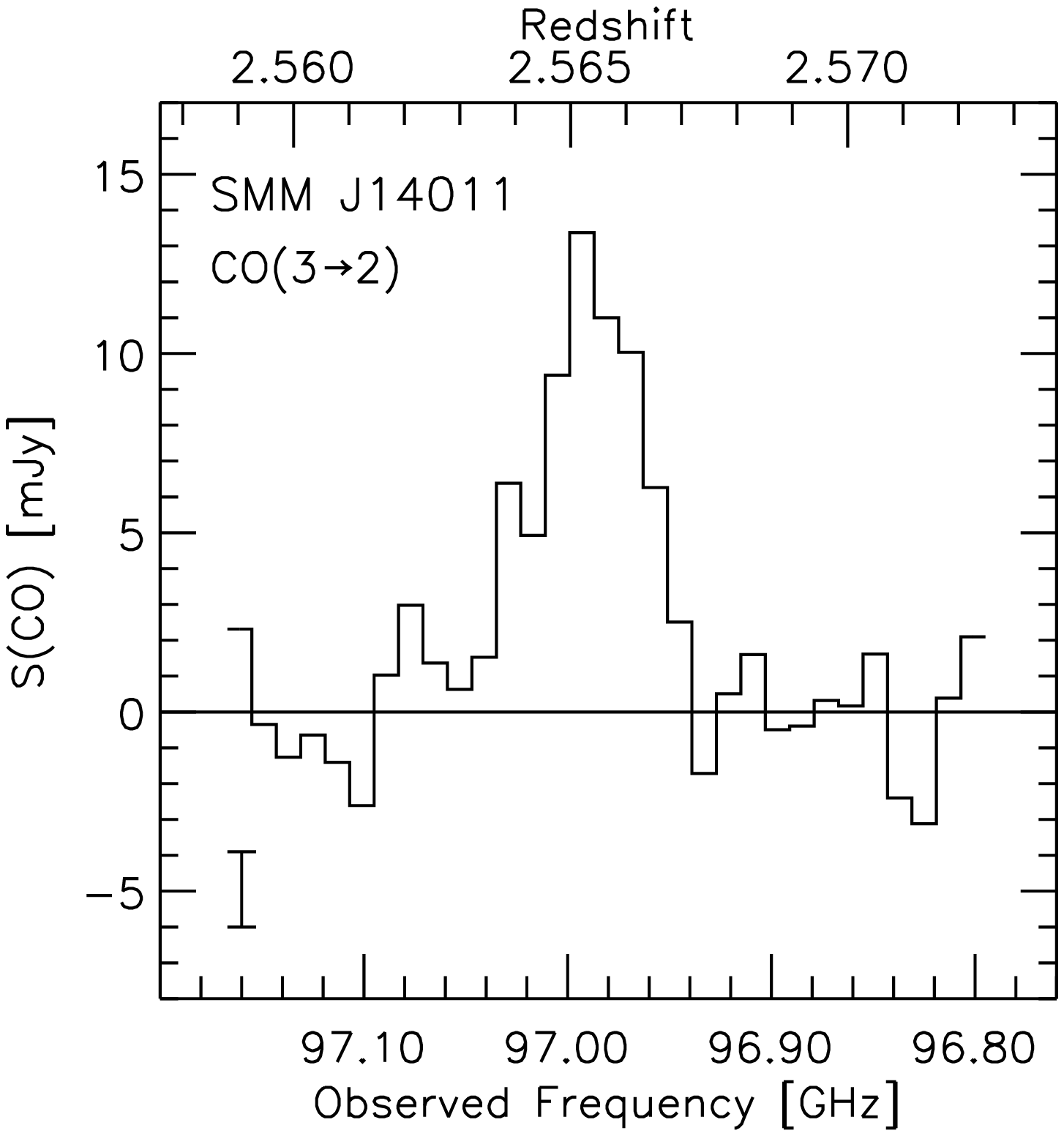]{The \coc spectrum for SMM\,J14011 observed with the OVRO
Millimeter Array.  The data have been smoothed to 24 MHz ($74\kps$) and
the channel separation is 12 MHz.  The $1\sigma$ rms error of each
channel is shown in the lower left.}

\figcaption[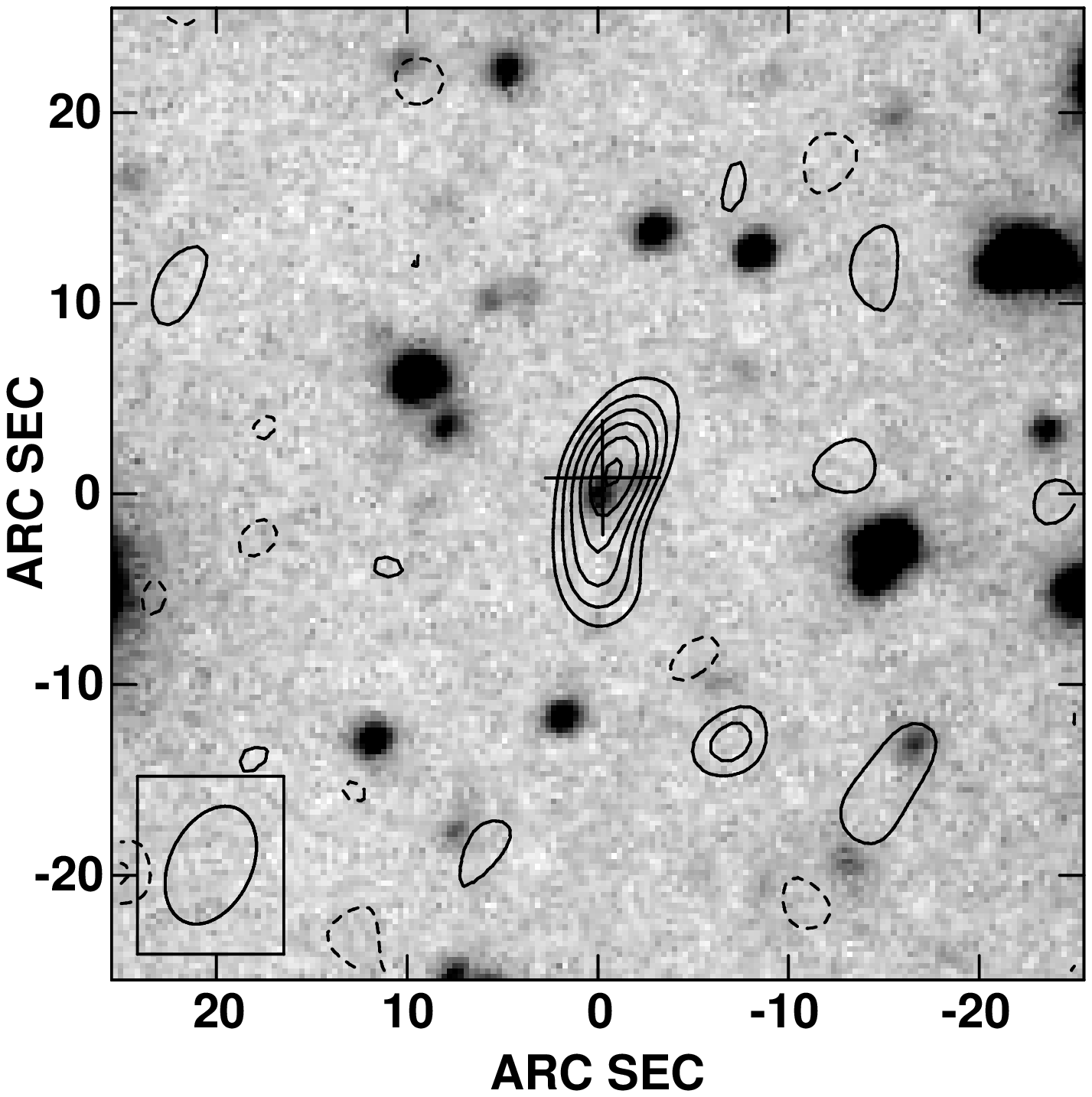]{The integrated \coc map averaged over 104 MHz
($322\kps$) overlaid on an optical $I$-band image taken at the Palomar
5\,m (Smail et al.\ 1998a).  The positional offsets are relative to the
optical position, and the cross marks the position of the SCUBA source
(Ivison et al.\ 1999).  The $1\sigma$ rms error is $0.32\jy\kps$/beam, and
the contour levels are $1\sigma\times$($-$3,$-$2,2,3,4,5,6,7).  The
synthesized beam size for the observations is shown in the lower left
($6\farcs5\times4\farcs3, {\rm PA}=-26\arcdeg$).}

\figcaption[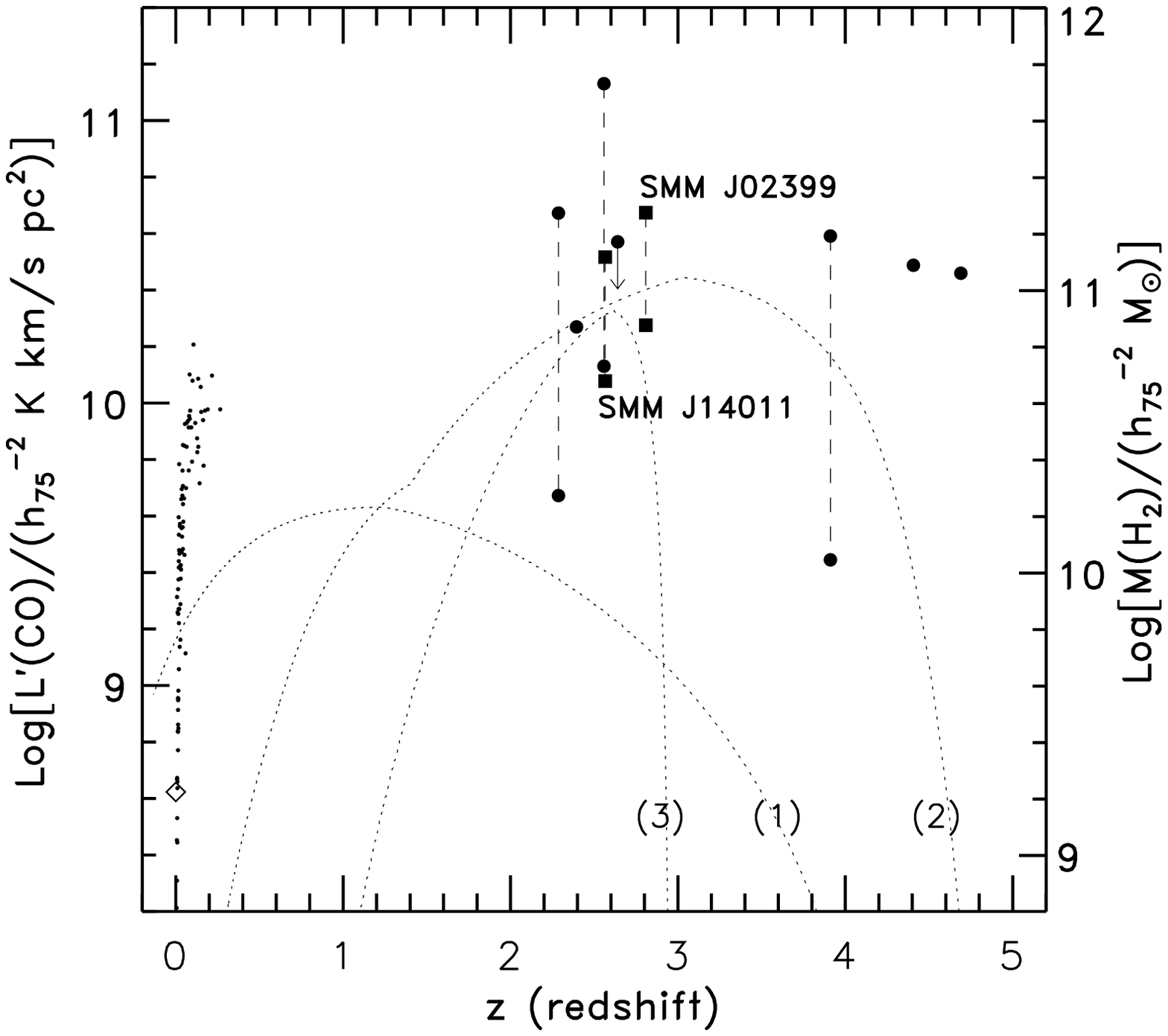]{The $L^{\prime}(\co)$ luminosity as a function of
redshift.  The molecular mass scale (at right) assumes
$\alpha=4\msun(\Kkpspc)^{-1}$.  Besides SMM\,J14011 and SMM\,J02399
(boxes), the other high--redshift CO sources (large dots) are IRAS
F\,10214+4724 at $z=2.3$ (Brown \& Vanden Bout 1991; Solomon et al.\
1992), 53W002 at $z=2.4$ (Scoville et al.\ 1997b), the $z=2.6$
Cloverleaf quasar H\,1413+117 (Barvainis et al.\ 1994), MG\,0414+0534 at
$z=2.6$ which has an unknown lensing magnification factor signified by
the downward arrow (Barvainis et al.\ 1998), APM\,08279+5255 at $z=3.9$
(Downes et al.\ 1998), BRI\,1335-0417 at $z=4.4$ (Guilloteau et al.\
1997), and BR\,1202-0725 at $z=4.7$ (Ohta et al.\ 1996; Omont et al.\
1996).  For known lensed sources, the observed and intrinsic values are
connected by a dashed line.  In addition, we plot (small dots) the low
redshift luminous IRAS sources from Sanders et al.\ (1991) and
additional ultraluminous IRAS sources from Solomon et al.\ (1997).  The
Milky Way Galaxy is shown as a diamond (Solomon \& Rivolo 1989).
Theoretical evolutionary curves are shown as dotted lines (see text):
(1) galactic disk model, (2) closed-box elliptical galaxy model, and (3)
starburst merger model.}

\begin{deluxetable}{ll}
\tablecaption{CO Observational Results}
\tablewidth{300pt}  
\tablehead{\colhead{Parameter}&\colhead{Value}}
\startdata
$\alpha$(J2000)  & $14^{\rm h} 01^{\rm m}04\fs92\pm0\fs03$ \nl
$\delta$(J2000) &$+02\arcdeg 52\arcmin 25\farcs6\pm0\farcs6$ \nl
$z(\co)$  &$2.5653\pm0.0003$ \nl
Linewidth (FWHM)& $200\pm40\kps$\nl
$S(\co)$\tablenotemark{a}& $2.4\pm0.3 \jy\kps$ \nl
$L(\co)$\tablenotemark{b}&  $1.6\times 10^{7} h_{75}^{-2}
L_{\sun}$ \nl 
$L^{\prime}(\co)$\tablenotemark{b}& $1.2\times 10^{10}
h_{75}^{-2}\Kkpspc$\nl     
$M(\hh)$\tablenotemark{b,c} & $5\times10^{10} h_{75}^{-2} M_{\sun}$\nl
\enddata 
\tablenotetext{a}{Observed \coc line flux.}
\tablenotetext{b}{Intrinsic value assuming a lensing amplification
factor of 2.75, $q_o =0.5$, and H$_o = 75\,h_{75} \kps \Mpc^{-1}$.}
\tablenotetext{c}{Estimated using $\alpha =4 \msun (\Kkpspc)^{-1}$.}
\end{deluxetable}

%\end{document}
%
%ApJL staff: ignore below (for preprints)
%
\setcounter{figure}{0}
\newpage
\begin{figure}
\includegraphics{f1.ps} \vspace*{7.0in}
\caption{ }
\end{figure}

\newpage
\begin{figure}
\includegraphics{f2.ps} \vspace*{7.0in}
\caption{ }
\end{figure}

\newpage
\begin{figure}
\includegraphics{f3.ps} \vspace*{7.in}
\caption{ }

\end{figure}

\end{document}